# $e^{\text{RPCA}}$: Robust Principal Component Analysis for Exponential Family Distributions

Xiaojun Zheng, Simon Mak, Liyan Xie, Yao Xie

*Abstract*—Robust Principal Component Analysis (RPCA) is a widely used method for recovering low-rank structure from data matrices corrupted by significant and sparse outliers. These corruptions may arise from occlusions, malicious tampering, or other causes for anomalies, and the joint identification of such corruptions with low-rank background is critical for process monitoring and diagnosis. However, existing RPCA methods and their extensions largely do not account for the underlying probabilistic distribution for the data matrices, which in many applications are known and can be highly non-Gaussian. We thus propose a new method called Robust Principal Component Analysis for Exponential Family distributions ($e^{\text{RPCA}}$), which can perform the desired decomposition into low-rank and sparse matrices when such a distribution falls within the exponential family. We present a novel alternating direction method of multiplier optimization algorithm for efficient $e^{\text{RPCA}}$ decomposition. The effectiveness of $e^{\text{RPCA}}$ is then demonstrated in two applications: the first for steel sheet defect detection, and the second for crime activity monitoring in the Atlanta metropolitan area.

## I. Introduction

With remarkable advances in sensing and experimental technologies, scientists and engineers now have access to massive datasets with complex forms for decision-making. The efficient harnessing of such data, in particular, the extraction of background structure and deviating anomalies (arising from occlusions, malicious tampering, process defects, or other causes for outliers), becomes ever more important for timely process monitoring, diagnosis and improvement. The increasingly complex form of such data further necessitates a careful consideration and integration of its underlying probabilistic distribution, which is often known and can be highly non-Gaussian. To tackle these challenges, we propose a novel Robust Principal Component Analysis method for Exponential Family Distributions ($e^{\text{RPCA}}$, for short), that leverages structure on this probabilistic distribution from the exponential family, to jointly perform anomaly detection and background extraction from massive and complex data matrices.

Our approach is motivated by two ongoing applications, on defect detection for steel sheet manufacturing and burglary monitoring in the Atlanta metropolitan area. For the first application, the timely detection of defects (e.g., gashes and dents) in steel sheet manufacturing is crucial for quality control. Recent developments in quanta image sensing (QIS; [11]) have shown promise for the desired high-frequency imaging, but these systems typically capture image intensities via binary bits. The efficient identification of potential defects for timely diagnosis thus poses a challenge with such large binary images. For the second application, the detection of regular and irregular burglary activities is paramount for crime monitoring and prevention. Reported burglary data naturally take the form of counts, and can be observed at high spatiotemporal resolution. Using such high-dimensional count data for timely extraction of regular and irregular crime patterns is thus a critical challenge. In both applications, both the identification of background structure, e.g., regular crime activity, and its corresponding anomalies, e.g., irregular crime activity, are important for timely decision-making from high-dimensional and complex data matrices.

A widely-used method for joint extraction of structure and sparse anomalies from a data matrix $\mathbf{M} \in \mathbb{R}^{p \times q}$ is the Robust Principal Component Analysis (RPCA) method [6]. RPCA decomposes $\mathbf{M}$ into the sum of two matrices $\mathbf{L}$ and $\mathbf{S}$, such that $\mathbf{L}$ is a low-rank matrix (modeling structure) and $\mathbf{S}$ is a sparse matrix (capturing anomalies). Such a decomposition can be optimized via convex optimization methods via its tightest convex relaxation using the nuclear and $l_1$-norms; more on this later. There has been much subsequent work on efficient optimization algorithms for the RPCA decomposition, including the use of augmented Lagrangian multipliers [27, 40], accelerated proximal gradient [2], alternating minimizing approaches [20] and low-rank matrix fitting [36]. There is also notable (albeit less) work on the RPCA when $\mathbf{M}$ is observed with random noise. This includes the stable principal component pursuit approach [42], which relaxes the equality constraint $\mathbf{M} = \mathbf{L} + \mathbf{S}$ to account for the presence of small measurement errors; more on this later. Such an approach, however, does not factor for the specific probabilistic distribution for $\mathbf{M}$, which in many applications may be known or can be reliably inferred. [13] proposed a hierarchical Bayesian approach for decomposing a noisy matrix into its low-rank and sparse components, but such an approach again does not factor in the non-Gaussian distribution of $\mathbf{M}$.

There is also a complementary line of work on extending the standard principal component analysis (PCA) for non-Gaussian noise distributions. This includes [12], which proposed a modification of PCA that generalizes to a broad class of so-called exponential family distributions [8] via Bregman distances. The exponential family covers a broad range of parametric distributions encountered in applications, including the Bernoulli, Poisson, Exponential, and Gaussian distribu-

XZ and SM are with the Department of Statistical Science, Duke University. LX is with the School of Data Science, Chinese University of Hong Kong (Shenzhen). YX is with the School of Industrial & Systems Engineering, Georgia Institute of Technology.

tions. [30] investigated a fully probabilistic extension of PCA for the exponential family. [28] presented the "Exponential PCA" (or ePCA) approach, which uses recent developments in random matrix theory and shrinkage for efficient estimation of low-rank structure under exponential family noise. Such methods, however, do not account for nor facilitate the identification of sparse anomalies in $\mathbf{M}$, which is critical in our aforementioned motivating applications.

To tackle these limitations, we thus propose a new $e^{\text{RPCA}}$ method that facilitates the *joint* extraction of low-rank structure and sparse anomalies, in the setting where data matrices are generated from the exponential family distribution. The $e^{\text{RPCA}}$ leverages a novel optimization formulation for this decomposition, which integrates information on the underlying probabilistic distribution of $\mathbf{M}$ via its likelihood function. We then present an alternating direction method of multiplier [3] optimization algorithm, which incorporates this distributional structure for efficient decomposition. Finally, we demonstrate the effectiveness of the $e^{\text{RPCA}}$ over existing methods in a suite of numerical experiments and for our two motivating applications on steel defect detection and crime monitoring. This suggests that when the distribution of the data matrices is either known or can be reliably inferred, the integration of such information can indeed be beneficial for a joint extraction of low-rank structure and its corresponding sparse anomalies.

This paper is organized as follows. Section II provides background on the RPCA, the ePCA, and their recent extensions, then discusses their limitations for our motivating application. Section III outlines the proposed $e^{\text{RPCA}}$, including its formulation and optimization algorithm, including a discussion on hyperparameter tuning and scalability for large data matrices. Section IV presents a suite of numerical experiments investigating the performance of $e^{\text{RPCA}}$ and existing methods, under different distributions of $\mathbf{M}$ from the exponential family. Section V explores the $e^{\text{RPCA}}$ in the aforementioned two motivating applications. Section VI concludes the paper.

## II. Background and Motivation

We first provide an overview of the robust PCA [6], the exponential PCA [28] and its extensions, then motivate the proposed $e^{\text{RPCA}}$ via our steel defect detection application.

### A. Robust PCA

RPCA [6] is a widely-used method for jointly recovering low-rank structure and anomalies from a data matrix $\mathbf{M} \in \mathbb{R}^{p \times q}$ with significant corruptions on a sparse number of entries. For recovering low-rank structure, it is well-known that the standard PCA approach [22, 35] can be highly sensitive to sparse and large outliers in $\mathbf{M}$; a single large outlier can greatly skew its estimated structure. PCA also cannot perform the task of detecting and isolating these sparse anomalies, which as mentioned before is critical for process diagnosis and quality control. To address such limitations, RPCA makes use of the following decomposition of the data matrix $\mathbf{M}$:

$$\min_{\mathbf{L},\mathbf{S}} \|\mathbf{L}\|_* + \lambda \|\mathbf{S}\|_1, \quad \text{s.t.} \quad \mathbf{L} + \mathbf{S} = \mathbf{M}. \tag{1}$$

Here, $\|\mathbf{A}\|_*$ is the nuclear norm (the sum of the singular values of $\mathbf{A}$), and $\|\mathbf{A}\|_1$ is the matrix $\ell_1$-norm (the sum of absolute values of entries in $\mathbf{A}$). Note that the nuclear norm $\mathbf{A}$ can be viewed as the tightest convex relaxation for the rank of $\mathbf{A}$, and its $\ell_1$-norm similarly serves as a convex relaxation of the number of non-zero entries in $\mathbf{A}$. Thus, Equation (1) decomposes the data matrix $\mathbf{M}$ as the sum of a low-rank matrix $\mathbf{L}$ and a sparse matrix $\mathbf{S}$, which facilitates the desired recovery of the underlying low-rank structures and anomalies from $\mathbf{M}$. The parameter $\lambda > 0$ controls the trade-off between low-rankedness and sparsity in this decomposition.

The formulation (1), which can be shown to be convex, can be efficiently optimized via a variety of scalable algorithms. A popular approach [27, 40] is to iteratively minimize the following augmented Lagrange multiplier (ALM) formulation:

$$\begin{aligned}\ell(\mathbf{L},\mathbf{S},\mathbf{Y}) :=& \|\mathbf{L}\|_* + \lambda \|\mathbf{S}\|_1 + \langle \mathbf{Y}, \mathbf{M} - \mathbf{L} - \mathbf{S}\rangle_F \\ &+ \frac{\mu}{2}\|\mathbf{M} - \mathbf{L} - \mathbf{S}\|_F^2,\end{aligned} \tag{2}$$

where $\langle \cdot, \cdot \rangle_F$ denotes the Frobenius inner product. Here, the equality constraint on $\mathbf{L} + \mathbf{S} = \mathbf{M}$ is replaced by its Lagrangian form, $\mathbf{Y}$ is the so-called Lagrange multiplier matrix, and $\mu > 0$ a positive constant. A generic Lagrange multiplier algorithm [1] can then be applied to iteratively solve (2). During the $k$-th iteration, one would optimize $(\mathbf{L}_k, \mathbf{S}_k) = \arg\min_{\mathbf{L},\mathbf{S}} \ell(\mathbf{L}, \mathbf{S}, \mathbf{Y}_k)$, then update the Lagrange multiplier matrix via $\mathbf{Y}_{k+1} = \mathbf{Y}_k + \mu(\mathbf{M} - \mathbf{L}_k - \mathbf{S}_k)$. These two steps are repeated until convergence. The parameter $\mu > 0$ can be viewed as the step size for updating the Lagrangian multiplier matrix. [27] provides further technical details on the validity and optimality of ALM. Further extensions include [38, 39], which explored various exact and inexact ALM approaches; [21], which proposed a linearized alternating direction optimization approach with adaptive penalties; and [2], which investigated the use of accelerated proximal gradient algorithms for performing this decomposition.

The formulation (1), however, does not account for the presence of noise in the observed data matrix $\mathbf{M}$, as $\mathbf{M}$ is assumed to decompose into the low-rank signal $\mathbf{L}$ and sparse anomalies $\mathbf{S}$ without noise. In many problems, including our motivating applications on steel defect detection and crime monitoring, such noise is ubiquitous and unavoidable; it arises either from the measurement process or as a realization of the data-generating process. To account for this, [42] investigated the following extension of RPCA, which they call the stable principal component pursuit (Stable-PCP):

$$\min_{\mathbf{L},\mathbf{S}} \|\mathbf{L}\|_* + \lambda \|\mathbf{S}\|_1 \quad \text{s.t.} \quad \|\mathbf{M} - \mathbf{L} - \mathbf{S}\|_F \le \delta. \tag{3}$$

The inequality in (3) relaxes the equality constraint $\mathbf{L} + \mathbf{S} = \mathbf{M}$ in (1) to account for a small amount of deviation $\delta$ resulting from noise. One limitation of the Stable-PCP, however, is that it does not factor in the *parametric form* of the underlying noise, which is often known in applications. For example, in imaging applications, the imaging system often dictates the parametric distribution used when observing the data matrix $\mathbf{M}$ [7]. As we shall see later, the use of this information on noise structure can greatly improve the recovery of both the low-rank signal $\mathbf{L}$ and its anomalies $\mathbf{S}$, particularly when such noise is large and non-Gaussian. We will investigate this further in later numerical experiments.

| Distribution | $\eta(\theta)$ | $t(m)$ | $a(\theta)$ | $b(m)$ |
|---|---|---|---|---|
| Poisson | $\log \theta$ | $m$ | $-\theta$ | $-\log(x!)$ |
| Bernoulli | $\log\left(\frac{\theta}{1-\theta}\right)$ | $m$ | $\log(1-\theta)$ | $0$ |
| Exponential | $-\theta$ | $m$ | $\log \theta$ | $0$ |
| Gaussian | $\frac{\theta}{\sigma}$ | $\frac{m}{\sigma}$ | $-\frac{\theta^2}{2\sigma^2}$ | $-\frac{\log(2\pi\sigma^2)}{2} - \frac{m^2}{2\sigma^2}$ |

TABLE I: Common distributions from the (one-parameter) exponential family (4): the Poisson, Bernoulli, Exponential, and Gaussian (with known variance $\sigma^2$) distributions.

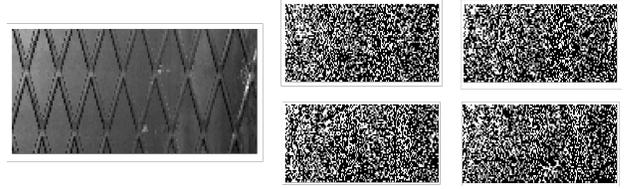

Fig. 1: (Left) The uncorrupted steel sheet image from Severstal, and (right) corresponding binary images generated with Bernoulli noise.

### B. Exponential PCA

The extension of the standard PCA for non-Gaussian noise has been explored in a series of papers; such work showed that when the parametric form of this noise is known, the integration of this structure can greatly improve the recovery of the low-rank signal. A common family of distributions is the exponential family distribution [8]. Given a single parameter $\theta$, the (one-parameter) exponential family is a family of distributions with probability density (or mass) function:

$$p_\theta(m) = \exp\left\{\eta(\theta)t(m) + a(\theta) + b(m)\right\}. \quad (4)$$

Here, $\eta(\theta)$ is the *canonical* parameterization of parameter $\theta$, $t(m)$ is the sufficient statistic of the distribution, and $a(\theta)$ and $b(m)$ are fixed and known functions of $\theta$ and data $m$, respectively. Such a specification defines a broad range of common distributions, including the Poisson, Bernoulli, Exponential, and Gaussian distributions (see Table I). We will denote a random variable following this distribution as $M \sim \text{ExpFam}\{\theta; \eta(\cdot), t(\cdot), a(\cdot), b(\cdot)\}$. An early work on integrating this non-Gaussian structure within PCA is [12], which proposed a PCA extension that generalizes to the exponential family distribution via Bregman distances. This is further extended in [30] via a fully probabilistic extension of PCA leveraging hybrid Monte Carlo sampling.

A recent development on this front is the ePCA method in [28]. The key idea is to leverage the eigendecomposition of a new covariance matrix estimator, constructed via moment calculations, shrinkage, and random matrix theory. ePCA begins with the sample covariance matrix of the data, then applies a series of operations, including diagonal debiasing, homogenization, shrinkage, heterogenization, and scaling (guided by the underlying exponential family model), to improve this covariance estimator. The corresponding low-rank representation can finally be obtained via an eigendecomposition of this modified covariance estimator. Further details of ePCA and its theoretical justification can be found in [28].

While such work on extending the standard PCA for non-Gaussian distributions is promising, there has been little work on leveraging such non-Gaussian noise for *jointly* recovering structure and anomalies in the presence of significant sparse outliers. This *combined* setting of non-Gaussian noise with sparse anomalies arises in a broad range of modern problems, including our two later applications on steel defect detection and crime monitoring. The aforementioned approaches, which tackle *only* the setting of non-Gaussian noise or sparse corruptions, can thus yield poor low-rank recovery and anomaly detection performance, as we will see next.

### C. Steel Defect Detection Application

We first investigate these existing methods for our motivating steel defect detection application (further details can be found in Section V-A). This application features the two defining challenges motivating our method: (i) non-Gaussian noise with (ii) significant sparse anomalies. For (i), the high-frequency imaging of steel sheets can be performed via quanta image sensing [11], which has shown improved performance over more conventional multi-bit systems (e.g., complementary metal-oxide semiconductor imaging [11]) due to higher frequency imaging with lower read noise [9]. QIS is a photon-counting device that captures image intensities using binary bits [11], which can be modeled via i.i.d. Bernoulli noise [9]. For (ii), anomalies arise in the form of defects in the steel manufacturing process, e.g., gashes, dents, or inhomogeneities on the steel sheet. These defects result in significant anomalies that are sparse on the imaged surface, and the primary objective is to quickly detect such anomalies for process diagnosis.

Fig. 1 (left) shows the uncorrupted image of a steel sheet from a steel industry company Severstal [34]. We see that the steel sheet has a "criss-cross" background structure, which can be well-represented via a low-rank decomposition. It also has visible defects from the manufacturing process, e.g., bumps and dents in white, particularly on the right side. To mimic QIS, we generate synthetic binary images by first normalizing the uncorrupted image intensities, then sampling $n = 500$ binary images from i.i.d. Bernoulli distributions with parameters taken as such intensities. Fig. 1 (right) shows several binary images generated in this fashion. We then explore the performance of RPCA and ePCA for jointly estimating the background structure of the steel sheet and its associated defects (further details on this set-up in Section V-A).

Fig. 2 shows the estimated low-rank structure $\mathbf{L}$ and its estimated anomalies (i.e., defects) $\mathbf{S}$ using RPCA, along with the estimated low-rank structure using ePCA. For RPCA, we see it yields a mediocre recovery of the criss-cross background, which is expected since it does not factor in the underlying *non-Gaussian* noise from QIS. Because of this, the estimated anomalies from RPCA erroneously capture the cross-cross background and fail to pinpoint the desired defects. The ePCA yields a slightly improved recovery of the background, but the recovered **L** also captures the underlying defects (in white), which is undesirable. This is again unsurprising since ePCA does not account for the presence of *sparse* anomalies. The combined setting of non-Gaussian noise with sparse anomalies thus poses a challenge for structure recovery and anomaly detection using these existing methods. We present next the proposed $e^{\text{RPCA}}$ approach for tackling these challenges.

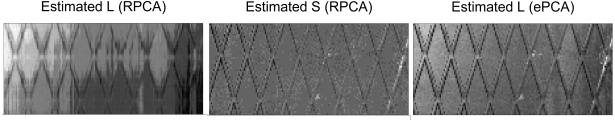

Fig. 2: Visualizing the estimated low-rank structure **L** from RPCA and its anomalies **S**, along with the estimated low-rank structure from ePCA, for the steel defect detection application.

## III. $e^{\text{RPCA}}$: RPCA for Exponential Family Distributions

We now outline the proposed Robust Principal Component Analysis for exponential family distributions ($e^{\text{RPCA}}$) for two settings: the "single-group" setting, where all observations share the *same* anomalies, and the "multi-group" setting, where anomalies may *change* between different groups of observations. We first formulate the optimization problem for the $e^{\text{RPCA}}$, then present an efficient optimization algorithm and recommendations for tuning parameters.

### A. Optimization Formulation

*1) Single-Group Setting:* Consider first the single-group setting. Suppose that we observe a collection of matrices $\mathbf{M}_1, \ldots, \mathbf{M}_n \in \mathbb{R}^{p \times q}$. Further suppose the entries of each $\mathbf{M}_i = (M_{i,j,k})_{j=1,\ldots,p;k=1,\ldots,q}$, follow the earlier exponential family model (4):

$$M_{i,j,k} \stackrel{indep.}{\sim} \text{ExpFam}(\theta_{j,k}; \eta(\cdot), t(\cdot), a(\cdot), b(\cdot)), \quad (5)$$

for $i = 1, \ldots, n$, $j = 1, \ldots, p$ and $k = 1, \ldots, q$. The random variable $M_{i,j,k}$ thus models the randomly corrupted observation given the true (unobserved) signal $\theta_{j,k}$. As before, we assume that prior knowledge is available on the class of noise distribution (e.g., Bernoulli), hence the functions $\eta(\cdot)$, $t(\cdot)$, $a(\cdot)$ and $b(\cdot)$ are known, and only the parameter matrix $\Theta = (\theta_{j,k})_{j=1,\ldots,p;k=1,\ldots,q}$ needs to be estimated.

Following RPCA, we assume the parameter matrix $\Theta$ can be decomposed as $\Theta = \mathbf{L} + \mathbf{S}$, where **L** is a low-rank matrix capturing background structure, and **S** is a sparse matrix that models for sparse anomalies. Let $l(\theta; m)$ be the negative log-likelihood function for the exponential family distribution (4) given a single data point $m$, defined as:

$$l(\theta; m) = -\eta(\theta)t(m) - a(\theta) - b(m). \quad (6)$$

One appeal of the exponential family is that, with the canonical parametrization $\eta(\theta)$ (or alternate careful parametrizations of $\theta$), the above negative log-likelihood can be made convex in the transformed parameter. This is important for our decomposition algorithm later; it allows for efficient parameter updates via computationally efficient convex optimization algorithms.

We can then formulate the optimization problem of the penalized maximum likelihood estimator [8] for $\Theta$ as:

$$\min_{\mathbf{L},\mathbf{S},\Theta} \sum_{i=1}^{n} \sum_{j=1}^{p} \sum_{k=1}^{q} \frac{l(\theta_{j,k}; M_{i,j,k})}{n} + \alpha \|\mathbf{L}\|_* + \beta \|\mathbf{S}\|_1 \quad (7)$$
$$\text{s.t.} \quad \Theta = \mathbf{L} + \mathbf{S}.$$

The first term in (7) is the standard maximum likelihood estimator for the parameter matrix $\Theta$. As in RPCA, the second term penalizes the rank of the background structure **L** via its tightest convex relaxation, and the third term penalizes the number of non-zero entries in **S** via its tightest convex relaxation. The parameters $\alpha > 0$ and $\beta > 0$ control the severity of each penalty term; Section III-C provides recommendations on how such parameters should be set.

Given that the noise corruption follows an exponential family distribution, one can then plug in the corresponding negative log-likelihood function $l$, and solve for **L** and **S** to extract the underlying low-rank structure and sparse anomalies. For example, in our steel defect application, where the image is subject to Bernoulli noise, the $e^{\text{RPCA}}$ formulation becomes:

$$\min_{\mathbf{L},\mathbf{S},\Theta} \sum_{i=1}^{n} \sum_{j=1}^{p} \sum_{k=1}^{q} \left\{ \frac{-M_{i,j,k} \log(\theta_{j,k})}{n} \right. \\ \left. - \frac{(1 - M_{i,j,k}) \log(1 - \theta_{j,k})}{n} \right\} + \alpha \|\mathbf{L}\|_* + \beta \|\mathbf{S}\|_1 \quad (8)$$
$$\text{s.t.} \quad \Theta = \mathbf{L} + \mathbf{S}.$$

Similar formulations can be adopted for other distributions from the exponential family (see Table I).

A natural question is whether the parameter matrix $\Theta$ itself is suitable for the desired low-rank plus sparse decomposition $\mathbf{L} + \mathbf{S}$, or whether such a decomposition is better suited on some transformation of $\Theta$. One computational advantage of the decomposition of $\Theta$ in (7) is that for the common distributions in Table I, one can show that the negative log-likelihood $l(\theta; m)$ is convex in $\theta$ (see [4]). As we shall see later, this convexity is useful for developing efficient optimization algorithms for solving the $e^{\text{RPCA}}$ formulation (7).

An alternate decomposition may be via its *canonical* parameterization $\eta(\theta)$ (see Table I). For example, in our steel defect application, suppose one expects the low-rank structure **L** to arise in the canonical parameter matrix, rather than $\Theta$. Defining the matrix $\mathbf{H} = (\eta_{j,k})_{j=1,\cdots,p;k=1,\cdots,q}$ as:

$$\eta_{j,k} = \log \frac{\theta_{j,k}}{1 - \theta_{j,k}}, \quad j = 1, \ldots, p, \ k = 1, \ldots, q,$$

the following formulation may be more appropriate:

$$\min_{\mathbf{L},\mathbf{S},\mathbf{E}} \sum_{i=1}^{n} \sum_{j=1}^{p} \sum_{k=1}^{q} \left\{ -\frac{M_{i,j,k} \eta_{j,k}}{n} + \frac{\log(1 + e^{\eta_{j,k}})}{n} \right\} \\ + \alpha \|\mathbf{L}\|_* + \beta \|\mathbf{S}\|_1 \quad (9)$$
$$\text{s.t.} \quad \mathbf{H} = \mathbf{L} + \mathbf{S}.$$

Similar formulations can be adopted for other exponential family distributions. This canonical decomposition again retains a convex formulation in the natural parameter $\eta$, and does not require additional constraints on **H** when solving (9), since the range of the canonical parameter $\eta$ is over the reals [8]. In problems where there is domain knowledge on where low-rank structure is expected to arise, such information should be used first and foremost for guiding the formulation of this low-rank plus sparse decomposition.

*2) Multi-Group Setting:* Consider next the multi-group setting, where between groups of observed matrices, the underlying low-rank structure remains the same but the sparse anomalies may change. This arises, e.g., for our crime monitoring application, where one may have common activities within each week, but different (and sparse) anomalies from week to week. Suppose the observed matrices form $G > 1$ groups, namely, $\{\mathbf{M}_i^{[1]}\}_{i=1}^{n_1}, \{\mathbf{M}_i^{[2]}\}_{i=1}^{n_2}, \ldots, \{\mathbf{M}_i^{[G]}\}_{i=1}^{n_G}$, where the anomalies may vary between different groups. As before, suppose the entries of each $\mathbf{M}_i^{[g]} = (M_{i,j,k}^{[g]})_{j=1,\ldots,p;k=1,\ldots,q}$ independently follow the exponential family distribution (4) with parameters in matrix $\Theta^{[g]}$.

With this, the multi-group $e^{\text{RPCA}}$ can be formulated as:

$$\min_{\mathbf{L},\mathbf{S}_1,\cdots,\mathbf{S}_G,\Theta} \sum_{j=1}^{p}\sum_{k=1}^{q}\left(\sum_{g=1}^{G}\sum_{i=1}^{n_g}\frac{l(\theta_{j,k}^{[g]};M_{i,j,k}^{[g]})}{n_g}\right) \\ + \alpha\|\mathbf{L}\|_* + \sum_{g=1}^{G}\beta_g\|\mathbf{S}_g\|_1 \quad (10) \\ \text{s.t.} \quad \Theta_g = \mathbf{L} + \mathbf{S}_g,\ g = 1,\ldots,G.$$

Here, each of the $G$ parameter matrices is decomposed as $\Theta_g = \mathbf{L} + \mathbf{S}_g$, where $\mathbf{L}$ models the common low-rank structure, and $\mathbf{S}_g$ models the different sparse anomalies within each group. The parameter $\alpha > 0$ controls the low-rank penalty, and the parameters $\beta_1,\ldots,\beta_G > 0$ control the sparsity within each group. One can similar adopt an alternate decomposition via the canonical parametrization $\eta(\theta)$ (see Table I).

### B. Optimization Algorithm

Next, we present efficient optimization algorithms for solving the single-group and multi-group $e^{\text{RPCA}}$ formulations (7) and (10). These algorithms make use of the alternating direction method of multiplier (ADMM) method [3], which has been widely used in large-scale optimization problems in image processing [10] and statistical learning [43]. ADMM optimizes problems of the form:

$$\min_{\mathbf{x},\mathbf{z}}\ f(\mathbf{x}) + g(\mathbf{z}) \quad \text{s.t.} \quad \mathbf{A}\mathbf{x} + \mathbf{B}\mathbf{z} = \mathbf{c}, \quad (11)$$

where $f$ and $g$ are convex functions of $\mathbf{x}$ and $\mathbf{z}$. Note that the original RPCA formulation (1) fits in the above form, where with $\mathbf{x} = \mathbf{L}$ and $\mathbf{z} = \mathbf{S}$, we have $f(\mathbf{x}) = \|\mathbf{L}\|_*$ and $g(\mathbf{z}) = \|\mathbf{S}\|_1$. With this, the key steps for ALM are to (i) minimize the augmented Lagrangian form (2) iteratively for first $\mathbf{L}$ then $\mathbf{S}$ (given other parameters fixed), (ii) update the Lagrange multiplier matrix via $\mathbf{Y}_{k+1} = \mathbf{Y}_k + \mu(\mathbf{M} - \mathbf{L}_k - \mathbf{S}_k)$. These steps are then iterated until the solution converges. One can show that this ADMM algorithm enjoys appealing convergence properties for RPCA optimization; see [27] for details. We will adapt ADMM for solving the single-group and multi-group problems.

*1) Single-Group Setting:* Consider first the single-group $e^{\text{RPCA}}$ problem (7). Its corresponding augmented Lagrangian form can be written as:

$$\min_{\mathbf{L},\mathbf{S},\Theta}\ \sum_{i=1}^{n}\sum_{j=1}^{p}\sum_{k=1}^{q}\frac{l(\theta_{j,k};M_{i,j,k})}{n} + \alpha\|\mathbf{L}\|_* + \beta\|\mathbf{S}\|_1 \\ + \langle \mathbf{Y}, \Theta - \mathbf{L} - \mathbf{S}\rangle_F + \frac{\mu}{2}\|\Theta - \mathbf{L} - \mathbf{S}\|_F^2, \quad (12)$$

where $\mathbf{Y}$ is the Lagrange multiplier matrix, and $\mu > 0$ is a constant. The key differences of the above $e^{\text{RPCA}}$ formulation from (2) are the additional parameter matrix $\Theta$ to optimize and the additional negative log-likelihood term in the objective.

Our optimization of (7) proceeds as follows. First, for fixed $\Theta$ and $\mathbf{L}$, the optimal $\mathbf{S}$ that minimizes (12) can be solved in closed form via the following lemma.

*Lemma 1 ([5]):* For $\tau > 0$, the optimal solution $\mathbf{S}^*$ to the following problem,

$$\min_{\mathbf{S}}\ \tau\|\mathbf{S}\|_1 + \frac{1}{2}\|\mathbf{X} - \mathbf{S}\|_F^2, \quad (13)$$

is given by:

$$S_{jk}^* = \mathcal{S}_\tau(\mathbf{X}) := \text{sgn}(X_{jk})\max(|X_{jk}| - \tau, 0), \quad (14)$$

for $j = 1,\ldots,p$ and $k = 1,\ldots,q$.

This is known as the *pointwise soft thresholding* solution. As a direct corollary, the optimal $\mathbf{S}$ that minimizes (12) given fixed $\Theta$ and $\mathbf{L}$, i.e.:

$$\mathbf{S}^* = \arg\min_{\mathbf{S}}\left\{\beta\|\mathbf{S}\|_1 + \frac{\mu}{2}\|\Theta - \mathbf{L} - \mathbf{S} + \frac{1}{\mu}\mathbf{Y}\|_F^2\right\}, \quad (15)$$

can be solved via pointwise soft thresholding (see Algorithm 1 for specific expression).

Similarly, for fixed $\Theta$ and $\mathbf{S}$, the optimal $\mathbf{L}$ that minimizes (12) can be solved in closed form via the lemma below.

*Lemma 2 ([5]):* Let $\mathbf{X} = \mathbf{U}\Sigma\mathbf{V}^T$ be the SVD of $\mathbf{X}$. Then for $\tau > 0$, the optimal solution $\mathbf{L}^*$ to the following problem

$$\arg\min_{\mathbf{L}}\ \tau\|\mathbf{L}\|_* + \frac{1}{2}\|\mathbf{X} - \mathbf{L}\|_F^2 \quad (16)$$

is given by:

$$\mathbf{L} = \mathcal{D}_\tau(\mathbf{X}) := \mathbf{U}\mathcal{S}_\tau(\Sigma)\mathbf{V}^T. \quad (17)$$

This is the *singular value thresholding* (SVT) solution. With this, the $\mathbf{L}$ that optimizes (12) given fixed $\Theta$ and $\mathbf{S}$:

$$\mathbf{L}^* = \arg\min_{\mathbf{L}}\left\{\alpha\|\mathbf{L}\|_* + \frac{\mu}{2}\|\Theta - \mathbf{L} - \mathbf{S} + \frac{1}{\mu}\mathbf{Y}\|_F^2\right\}, \quad (18)$$

can be solved via SVT (see Algorithm 1 for expression).

Finally, we need to optimize (12) for the parameter matrix $\Theta$ given $\mathbf{L}$ and $\mathbf{S}$. Note that this can be *decoupled* into $pq$ separate optimization problems for each entry of $\Theta$, i.e., for $j = 1,\ldots,p$ and $k = 1,\ldots,q$:

$$\theta_{j,k}^* = \arg\min_{\theta_{j,k}}\left\{\sum_{i=1}^{n}\frac{l(\theta_{j,k};M_{i,j,k})}{n} \right. \\ \left. + \frac{\mu}{2}\left(\theta_{j,k} - L_{j,k} - S_{j,k} + \frac{1}{\mu}Y_{j,k}\right)^2\right\} \quad (19) \\ =: \arg\min_{\theta_{j,k}}\zeta_{j,k}(\theta_{j,k};L_{j,k},S_{j,k},Y_{j,k}).$$

For the exponential family, it is known that the negative log-likelihood $l(\theta; m)$ is *convex* in $\theta$ for the common distributions in Table I (see [4]). Thus, given $\mathbf{L}$ and $\mathbf{S}$, we can optimize for $\Theta$ using gradient descent methods and enjoy standard convergence guarantees [31, 33]. For certain exponential family distributions (e.g., the Bernoulli), one can further obtain closed-form solutions for (19) that can be exploited for efficient optimization (see Appendix for such closed-form solutions for specific distributions). The decoupled problem (19) can further be sped up via parallel optimization on each entry of $\Theta$.

With this in hand, the proposed optimization algorithm is presented in Algorithm 1. We begin with an initial estimate on $\mathbf{L}$ via a low-rank SVD approximation of the observation mean $\bar{\mathbf{M}} = (1/n)\sum_{i=1}^n \mathbf{M}_i$, with $\mathbf{Y}$ initialized at $\mathbf{0}$ and $\Theta$ initialized at $\bar{\mathbf{M}}$. Next, we update $\mathbf{L}$ as the singular value thresholding solution to (18) given current iterates for $\mathbf{S}$, $\mathbf{Y}$ and $\Theta$, then update $\mathbf{S}$ as the soft thresholding solution to (15) given current $\mathbf{L}$, $\mathbf{Y}$ and $\Theta$. The parameter matrix $\Theta$ is then optimized via (19) either gradient descent methods [33] (the L-BFGS algorithm [29] was used in later experiments) or the closed-form updates in Table II (see Appendix). Finally, the Lagrange multiplier matrix $\mathbf{Y}$ is updated via:

$$\mathbf{Y} \leftarrow \mathbf{Y} + \mu(\Theta - \mathbf{L} - \mathbf{S}), \tag{20}$$

where $\mu > 0$ is a step size parameter. These steps are then iterated until convergence. Algorithm 1 summarizes the detailed steps of this optimization procedure.

It is worth noting that, in Algorithm 1, the step size $\mu$ is fixed as a constant, i.e., $\mu_t = \mu$ over different iterations $t$. This can be justified as follows. For the ALM formulation of RPCA (2), one can show that (see [27]) if $(\mu_t)_{t=1}^\infty$ is a non-decreasing sequence and $\sum_{t=1}^\infty \mu_t^{-1} = \infty$, then the iteratively updated matrices $\mathbf{L}^{[t]}$ and $\mathbf{S}^{[t]}$ converge to an optimal solution $(\mathbf{L}^*, \mathbf{S}^*)$ for the RPCA problem (1). Furthermore, if $\mu_t$ is bounded above, one can show that the iterative solutions $(\mathbf{L}^{[t]}, \mathbf{S}^{[t]})$ can reach $\epsilon$-optimality from the optimal solution (i.e., within $\epsilon$ of the desired RPCA objective in (1)) after $t = \mathcal{O}(1/\epsilon)$ algorithm iterations [27]. Given such convergence guarantees for the RPCA, we thus adopt a similar strategy of constant step size $\mu$ for the $e^{\text{RPCA}}$ via Algorithm 1. We further note that, while theoretical convergence guarantees are difficult to establish for Algorithm 1 (due in large part to the iterative estimation of the unknown natural parameter matrix), empirical experiments later suggest that the employed algorithm yields satisfactory convergence and optimization performance.

*2) Multi-Group Setting:* Consider next the multi-group $e^{\text{RPCA}}$ problem (10), where there are multiple groups of observed matrices $\{\mathbf{M}_i^{[1]}\}_{i=1}^{n_1}, \{\mathbf{M}_i^{[2]}\}_{i=1}^{n_2}, \ldots, \{\mathbf{M}_i^{[G]}\}_{i=1}^{n_G}$ that share a common low-rank structure $\mathbf{L}$ but different sparse anomalies $\mathbf{S}_1, \ldots, \mathbf{S}_G$. We adopt a similar augmented Lagrangian form for optimization, given by:

$$\min_{\mathbf{L},\mathbf{S}_1,\ldots,\mathbf{S}_G,\Theta} \sum_{j=1}^p \sum_{k=1}^q \left( \sum_{g=1}^G \sum_{i=1}^{n_g} \frac{l(\theta_{j,k}^{[g]}; M_{i,j,k}^{[g]})}{n_g} \right) + \alpha \|\mathbf{L}\|_*$$

$$+ \sum_{g=1}^G \left\{ \beta_g \|\mathbf{S}_g\|_1 + \langle \mathbf{Y}_g, \Theta_g - \mathbf{L} - \mathbf{S}_g \rangle_F \right. \tag{21}$$

$$\left. + \frac{\mu}{2} \|\Theta_g - \mathbf{L} - \mathbf{S}_g\|_F^2 \right\}.$$

---

**Algorithm 1** Single-group $e^{\text{RPCA}}$ optimization via ADMM

**Inputs**: Data matrices $\mathbf{M}_1, \ldots, \mathbf{M}_n$, initial parameters $(\mathbf{S}^{[0]}, \mathbf{Y}^{[0]}, \Theta^{[0]})$, penalty parameters $\alpha, \beta, \mu$.
**Initialize** $\mathbf{S} = \mathbf{S}^{[0]}$, $\mathbf{Y} = \mathbf{Y}^{[0]}$ and $\Theta = \Theta^{[0]}$. Set $t = 0$.
**while** not converge **do**
$\quad \mathbf{L}^{[t+1]} \leftarrow \mathcal{D}_{\alpha/\mu}(\Theta^{[t]} - \mathbf{S}^{[t]} + \frac{1}{\mu}\mathbf{Y}^{[t]})$
$\quad \mathbf{S}^{[t+1]} \leftarrow \mathcal{S}_{\beta/\mu}(\Theta^{[t]} - \mathbf{L}^{[t+1]} + \frac{1}{\mu}\mathbf{Y}^{[t]})$
$\quad$ **for** $j = 1, \ldots, p$ and $k = 1, \ldots, q$
$\quad\quad \theta_{j,k}^{[t+1]} \leftarrow \arg\min_{\theta_{j,k}} \zeta_{j,k}(\theta_{j,k}; L_{j,k}^{[t+1]}, S_{j,k}^{[t+1]}, Y_{j,k}^{[t]})$
$\quad$ **end for**
$\quad \mathbf{Y}^{[t+1]} \leftarrow \mathbf{Y}^{[t]} + \mu(\Theta^{[t+1]} - \mathbf{L}^{[t+1]} - \mathbf{S}^{[t+1]})$
$\quad$ Update $t \leftarrow t + 1$.
**end while**
**Outputs**: Optimized parameters $(\mathbf{S}^{[t]}, \mathbf{L}^{[t]}, \Theta^{[t]})$.

---

Similar to before, $\mathbf{Y}_1, \ldots, \mathbf{Y}_G$ are Lagrange multiplier matrices, and $\mu > 0$ is a constant.

The optimization problem (21), unfortunately, is harder to solve than the single-group problem (12). This is due to the fact that, as the low-rank structure $\mathbf{L}$ is shared over all groups, its optimization, given fixed $\Theta, \mathbf{S}_1, \ldots, \mathbf{S}_G$, is no longer in closed form. We thus adopt the following two-stage algorithm to find an approximate solution. For Stage 1, we obtain an *estimate* $\widetilde{\mathbf{L}}$ of the low-rank matrix $\mathbf{L}$ under the approximation $\mathbf{S} = \mathbf{S}_1 = \cdots = \mathbf{S}_G$, i.e., all groups have the same sparse anomalies, and thus $\Theta = \Theta_1 = \cdots = \Theta_G$. This can be achieved via a direct application of the earlier single-stage ADMM algorithm. The optimization of $\widetilde{\mathbf{L}}$ (given common parameters $\Theta$ and common anomalies $\mathbf{S}$) yields the closed-form singular value thresholding update:

$$\widetilde{\mathbf{L}}^{[t]} \leftarrow \mathcal{D}_{\alpha/\mu}\left(\Theta^{[t]} - \mathbf{S}^{[t]} + \frac{1}{\mu}\mathbf{Y}^{[t]}\right). \tag{22}$$

Similarly, the optimization of $\mathbf{S}$ (given $\widetilde{\mathbf{L}}$ and $\Theta$) and $\Theta$ (given $\widetilde{\mathbf{L}}$ and $\mathbf{S}$) yields closed-form updates from (15) and (19), respectively. For Stage 2, with $\mathbf{L}$ *fixed* at this estimated $\widetilde{\mathbf{L}}$, we then cyclically optimize the group-dependent sparse anomaly matrices $\mathbf{S}_1, \ldots, \mathbf{S}_G$ and parameter matrices $\Theta_1, \ldots, \Theta_G$ via a similar ADMM algorithm on (21); such closed-form updates are provided in Algorithm 2. The detailed steps for this two-stage optimization procedure are outlined in Algorithm 2.

### C. Hyperparameter Tuning

Finally, careful tuning of the hyperparameters $\alpha$, $\beta$, and $\mu$ is needed for accurate recovery of the low-rank structure and sparse anomalies. For the standard RPCA, [6] showed that with $\alpha = 1$, $\beta = 1/\sqrt{\max(p,q)}$ and $\mu = pq/(4\|\Theta\|_1)$, one achieves the theoretical recovery of $\mathbf{L}$ and $\mathbf{S}$ in an asymptotic sense. We found that such a specification works reasonably well for the $e^{\text{RPCA}}$ as well for both single-group and multi-group settings, in the absence of any prior knowledge on the rank of $\mathbf{L}$ or the degree of sparsity in $\mathbf{S}$. We note, however, that this specification presumes no noise in the observation of $\mathbf{L}+\mathbf{S}$; when large non-Gaussian noise is present, we found that a larger choice of $\mu$ may yield improved recovery performance.

7**Algorithm 2** Multi-group $e^{\text{RPCA}}$ optimization via ADMM

*Inputs*: Data matrices $\{\mathbf{M}_i^{[1]}\}_{i=1}^{n_1}, \ldots, \{\mathbf{M}_i^{[G]}\}_{i=1}^{n_G}$, initial parameters for Stage 1 $\{(\mathbf{S}^{[0]}, \mathbf{Y}^{[0]}, \Theta^{[0]})\}$ and Stage 2 $\{(\mathbf{S}_g^{[0]}, \mathbf{Y}_g^{[0]}, \Theta_g^{[0]})\}_{g=1}^G$, penalty parameters $\alpha, \beta, \mu$.

*Stage 1*: Let $\widetilde{\mathbf{L}}$ be the low-rank structure with common anomalies $\mathbf{S}$ and parameters $\Theta$.
**Initialize** $\mathbf{S} = \mathbf{S}^{[0]}$, $\mathbf{Y} = \mathbf{Y}^{[0]}$ and $\Theta = \Theta^{[0]}$.
**Optimize** $\widetilde{\mathbf{L}}$ using Algorithm 1.

*Stage 2*: Fix $\mathbf{L} = \widetilde{\mathbf{L}}$.
**for** $g = 1, \ldots, G$
  **Initialize** $\mathbf{S}_g = \mathbf{S}_g^{[0]}, \mathbf{Y}_g = \mathbf{Y}_g^{[0]}, \Theta_g = \Theta_g^{[0]}$. Set $t = 0$.
  **while** not converge **do**
    $\mathbf{S}_g^{[t+1]} \leftarrow \mathcal{S}_{\beta_g/\mu}(\Theta_g^{[t]} - \mathbf{L} + \frac{1}{\mu}\mathbf{Y}_g^{[t]})$
    **for** $j = 1, \ldots, p$ and $k = 1, \ldots, q$
      $\theta_{g,j,k}^{[t+1]} \leftarrow \arg\min_{\theta_{g,j,k}} \zeta_{g,j,k}(\theta_{g,j,k}; L_{j,k}, S_{g,j,k}^{[t+1]}, Y_{g,j,k}^{[t]})$
    **end for**
    $\mathbf{Y}_g^{[t+1]} = \mathbf{Y}_g^{[t]} + \mu(\Theta_g^{[t+1]} - \mathbf{L} - \mathbf{S}_g^{[t+1]})$
    Update $t \leftarrow t + 1$.
  **end while**
**end for**
**Outputs**: Optimized parameters $\{(\mathbf{S}_g^{[t]}, \Theta_g^{[t]})\}_{g=1}^G, \mathbf{L}$.

---

**Algorithm 3** Hyperparameter tuning for single-group $e^{\text{RPCA}}$

*Inputs*: Hyperparameter step sizes $\eta_\alpha > 0, \eta_\beta > 0$.
*Condition* $(*)$: $\text{rank}(\mathbf{L}) \leq r$, $\%\text{nz}(\mathbf{S}) < s$

**Initialize** $\alpha^{[0]} = 1$, $\beta^{[0]} = 1/\sqrt{\max(p,q)}$. Set $t = 0$.
**Optimize** $(\mathbf{L}, \mathbf{S})$ with $\alpha = \alpha^{[0]}, \beta = \beta^{[0]}$ via Algorithm 2.
**While** $(*)$ is not satisfied **do**
  **if** $\text{rank}(\mathbf{L}) > r$ **then**
    $\alpha^{[t]} \leftarrow \alpha^{[t-1]} + \eta_\alpha \sqrt{t}$
  **else** $\alpha^{[t]} \leftarrow \alpha^{[t-1]}$
  **end if**
  **if** $\%\text{nz}(\mathbf{S}) > s$ **then**
    $\beta^{[t]} \leftarrow \beta^{[t-1]} + \eta_\beta \sqrt{t}$
  **else** $\beta^{[t]} \leftarrow \beta^{[t-1]}$
  **end if**
  **Optimize** $(\mathbf{L}, \mathbf{S})$ with $\alpha = \alpha^{[t]}, \beta = \beta^{[t]}$ via Algorithm 2.
  Update $t \leftarrow t + 1$.
  Stop if $\alpha^{[t]} = \alpha^{[t-1]}$ and $\beta^{[t]} = \beta^{[t-1]}$.
**end while**
**Outputs**: Optimized hyperparameters $(\alpha^{[t]}, \beta^{[t]})$.

---

In many applications, there may be guiding prior information on the rank of $\mathbf{L}$ or the sparsity level of $\mathbf{S}$, which can integrated for hyperparameter tuning. We illustrate how this can be done for the single-group setting; a similar strategy can be used for multiple groups. Suppose we have a desired upper bound on the rank $\text{rank}(\mathbf{L}) \leq r$ and the proportion of non-zero entries $\%\text{nz}(\mathbf{S}) := \#\{S_{j,k} \neq 0\}/(pq) < s$. Starting with an initial hyperparameter setting $\alpha^{[0]} = 1$, $\beta^{[0]} = 1/\sqrt{\max(p,q)}$ and $\mu = pq/(4\|\Theta\|_1)$, we first perform the $e^{\text{RPCA}}$ optimization (Algorithm 1) with $\alpha = \alpha^{[0]}$ and $\beta = \beta^{[0]}$ to estimate the low-rank structure $\mathbf{L}$ and corresponding anomalies $\mathbf{S}$. If the rank of the estimated $\mathbf{L}$ exceeds $r$, we then iteratively increase $\alpha$. Otherwise, if the proportion of sparse entries in the estimated $\mathbf{S}$ exceeds $s$, we then iteratively increase $\beta$. Algorithm 3 summarizes this hyperparameter tuning procedure.

*D. Computational Complexity*

Another appealing property of the $e^{\text{RPCA}}$ is that, in addition to leveraging the underlying exponential family structure, it also permits closed-form efficient updates in the optimization algorithm. We investigate next the computational complexity of $e^{\text{RPCA}}$ optimization algorithm, in terms of the data matrix dimensions $p$ and $q$ as well as sample size $n$.

For the single-group setting (Algorithm 1), each iterative update of $\mathbf{L}$ involves an SVD operation that requires $\mathcal{O}(pq\min(p,q))$ work. Each iterative update of $\mathbf{S}$ and $\mathbf{Y}$ requires $\mathcal{O}(pq)$ work. Each iterative update for the parameter $\theta_{j,k}$ (of which there are $pq$ in total) requires $\mathcal{O}(1)$ work, but this involves a one-shot computation of the sample mean $\bar{\mathbf{M}} = n^{-1}\sum_{i=1}^n \mathbf{M}_i$ (see the closed-form updates in the Appendix) that requires $\mathcal{O}(npq)$ work. Summarizing the above, the single-group Algorithm 1 thus requires an initial $\mathcal{O}(npq)$ work for pre-processing, and $\mathcal{O}(pq\min(p,q))$ work per optimization iteration. For the multi-group setting (Algorithm 2), we need to consider both Stage 1 and Stage 2. For Stage 1, one can follow the above rationale to show that this requires an initial $\mathcal{O}(npq)$ work for pre-processing, and $\mathcal{O}(pq\min(p,q))$ work per optimization iteration. For Stage 2, the parameter updates within each group require $\mathcal{O}(pq)$ work, thus incurring a total work of $\mathcal{O}(Gpq)$ per optimization iteration.

With a relatively small sample size $n$ and number of groups $G$, the key computational bottleneck for $e^{\text{RPCA}}$ thus lies in the SVT updates that each require $\mathcal{O}(pq\min(p,q))$ work. This will not be too burdensome for data matrices with large $p$ and small $q$ (or vice versa), but may be time-consuming when both $p$ and $q$ are large. Luckily, with modern computing architecture, such operations can be greatly sped up via multi-thread processing and GPU acceleration; see, e.g., [16, 25]. As this is the primary bottleneck for $e^{\text{RPCA}}$, we have found such tools to be greatly useful in scaling up this decomposition approach for massive data matrices.

We also provide next a brief comparison of computational complexity with the standard RPCA and the ePCA. For the standard RPCA (1) (with appropriate modifications for the considered noisy setting with multiple data matrices; see Section IV for details), the ALM approach requires the same running time per optimization iteration of $\mathcal{O}(pq\min(p,q))$. The ePCA incurs much higher computation relative to the other two methods, particularly for large data matrices. The key computational bottleneck in ePCA is the eigendecomposition of a $pq \times pq$ covariance matrix, which requires $\mathcal{O}((pq)^3)$ work. Clearly, for massive data matrices with both $p$ and $q$ large, the $e^{\text{RPCA}}$ and standard RPCA are much more computationally efficient compared to the ePCA.

IV. NUMERICAL EXPERIMENTS

We now explore the performance of the proposed $e^{\text{RPCA}}$ in a suite of simulation experiments. Here, the underlying parameter matrix $\Theta \in \mathbb{R}^{p \times q}$ follows the presumed low-rank plus sparse decomposition $\Theta = \mathbf{L} + \mathbf{S}$. The low-rank matrix $\mathbf{L}$ is simulated by first generating a matrix with independent entries from the Gaussian distribution $\mathcal{N}(\zeta, \gamma^2)$, then truncating all but the largest $p/5$ singular values via SVD. The sparse matrix $\mathbf{S}$ is simulated with $p^2/20$ uniformly selected non-zero entries, where non-zero entry values are uniformly sampled from the interval $[L, U]$. The corresponding data matrices are then generated from $\Theta$ following several common non-Gaussian exponential family distributions, including the Bernoulli, Exponential, and Poisson distributions. Since the entries of $\Theta$ should be constrained to specific intervals depending on the choice of exponential family distribution, the simulation parameters $\zeta, \gamma^2, L, U$ need to be carefully chosen to adhere to such constraints; for brevity, we provided specific settings of these parameters in the Appendix.

For comparison, we adopt two baseline approaches. The first is the standard RPCA approach [6]. Since *multiple* data matrices $\mathbf{M}_1, \ldots, \mathbf{M}_n$ are observed with *noise*, the RPCA formulation (1) can be modified as follows for fair comparison:

$$\min_{\mathbf{L},\mathbf{S},\Theta} \sum_{i=1}^{n} \sum_{j=1}^{p} \sum_{k=1}^{q} \frac{(M_{i,j,k} - \theta_{j,k})^2}{2n\hat{\sigma}^2} + \|\mathbf{L}\|_* + \lambda \|\mathbf{S}\|_1 \quad (23)$$
$$\text{s.t.} \quad \Theta = \mathbf{L} + \mathbf{S},$$

where $\hat{\sigma}$ is the sample standard deviation of the data matrices, and $\lambda$ is set via recommended settings in [6]. The formulation (23) makes two modifications in its first term, which account for standard Gaussian noise as well as multiple data matrices $\mathbf{M}_1, \ldots, \mathbf{M}_n$. This is analogous to the Stable-PCP formulation (3) from [42] (for which we could not find an implementation online), where the constraint in (3) is relaxed via the first term in (23). This is then optimized using the aforementioned ALM approach [27, 40] (see Section II-A), with the recommended step size $\mu = pq/(4\|\hat{\Theta}\|_1)$ in [6], where $\hat{\Theta}$ is the maximum likelihood estimator [8] of $\Theta$. The second baseline is the ePCA method [28], which can leverage the underlying non-Gaussian noise for extracting low-rank structure, but not for detection of sparse anomalies. These methods are compared with the $e^{\text{RPCA}}$ for the single-group and multi-group settings.

## A. Single-Group Setting

In the single-group experiments, we generated the low-rank matrix $\mathbf{L}$ and sparse anomalies $\mathbf{S}$ as described previously, then simulated $n = 500$ data matrices $\mathbf{M}_1, \ldots, \mathbf{M}_n$ from the parameter matrix $\Theta = \mathbf{L} + \mathbf{S}$. We then compared each method on its recovery of the low-rank structure $\mathbf{L}$ and the sparse anomalies $\mathbf{S}$ in terms of Frobenius error. These experiments are performed for matrices of dimensions $p \times p$, $p = 10, 20, 30$ and $40$, with each setting replicated for 30 trials.

*1) Bernoulli Distribution:* Consider first noise drawn from the Bernoulli distribution, where each matrix entry follows $M_{i,j,k} \stackrel{indep.}{\sim} \text{Bern}(\theta_{j,k})$. Fig. 3a shows the recovery errors for $\mathbf{L}$ and $\mathbf{S}$ (in Frobenius norm) using the proposed $e^{\text{RPCA}}$, the standard RPCA (with appropriate modifications detailed earlier) and the ePCA; note that the latter does not provide an estimate of $\mathbf{S}$. For $\mathbf{L}$, we see that the $e^{\text{RPCA}}$ yields a noticeably improved recovery of the low-rank structure compared to existing methods, particularly as dimension $p$ increases. Similarly, for $\mathbf{S}$, we observe an improved recovery of the underlying sparse anomalies for the $e^{\text{RPCA}}$ compared to RPCA, with this improvement again growing as dimension $p$ increases. This suggests, that with prior information on the underlying non-Gaussian noise, the integration of such information can greatly improve the joint recovery of both the underlying low-rank structure and sparse anomalies.

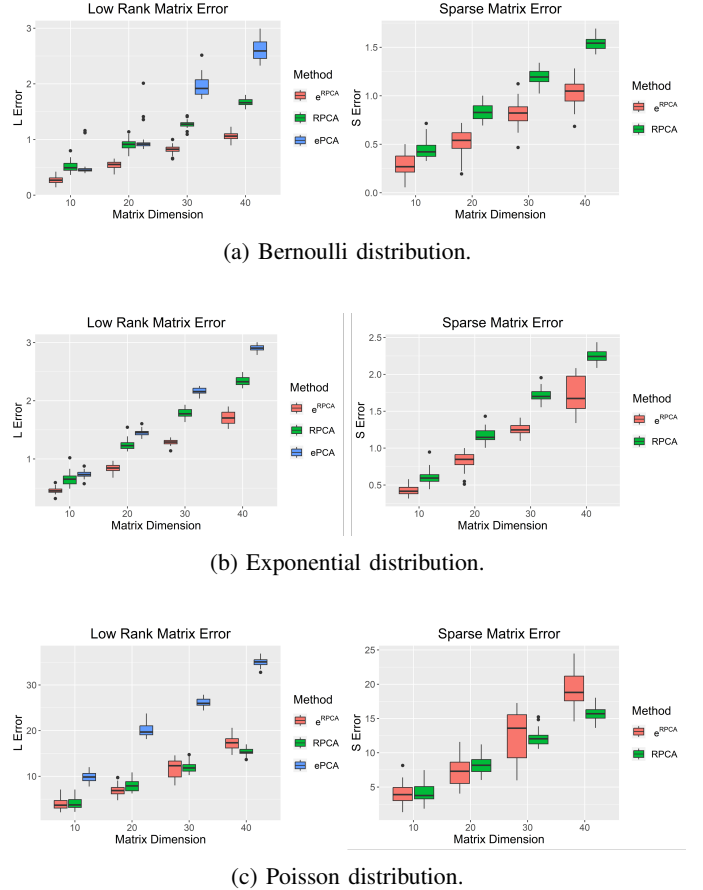

(a) Bernoulli distribution.

(b) Exponential distribution.

(c) Poisson distribution.

Fig. 3: Boxplots of recovery errors (in Frobenius norm) for the low-rank matrix $\mathbf{L}$ (left), and sparse anomalies $\mathbf{S}$ (right) as a function of matrix dimension $p$ for single-group simulations across different distributions.

Figs. 4a and 4b visualize the recovery of $\mathbf{L}$ and $\mathbf{S}$, respectively, for one simulation experiment in $p = 10$ dimensions. We see that the standard RPCA yields a noticeably poorer recovery of the true $\mathbf{L}$ compared to $e^{\text{RPCA}}$, which is unsurprising as it does not account for the underlying non-Gaussian noise form. This, in turn, resulted in the erroneous detection of many "anomalies" that were not truly anomalies. The ePCA, while providing slightly better recovery of $\mathbf{L}$ to the RPCA, can be seen to be highly sensitive to the underlying sparse anomalies, resulting in significant deterioration of the recovery. The proposed $e^{\text{RPCA}}$, by incorporating the underlying non-Gaussian noise within the desired low-rank plus sparse decomposition, yields improved recovery of both the low-rank structure $\mathbf{L}$ and the sparse anomalies $\mathbf{S}$.

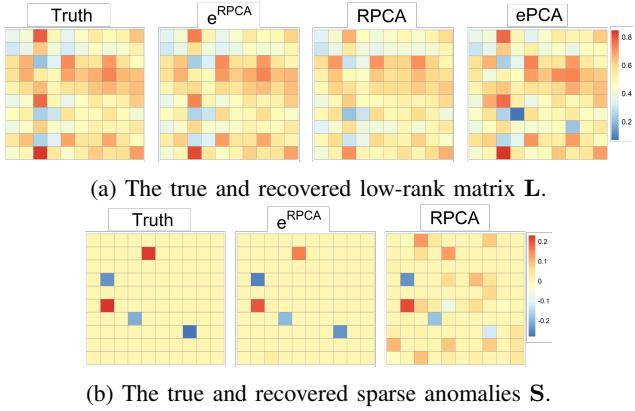

(a) The true and recovered low-rank matrix **L**.

(b) The true and recovered sparse anomalies **S**.

Fig. 4: Visualizing the true and recovered low-rank matrix **L** and sparse anomalies **S** in one simulation in $p = 10$ dimensions, for the single-group Bernoulli simulations.

*2) Exponential Distribution:* Consider the case where noise is drawn from the Exponential distribution; thus each entry follows $M_{i,j,k} \overset{indep.}{\sim} \text{Exp}(\theta_{j,k})$ and $\theta_{j,k}$ is its rate parameter. Here, since the sparse decomposition arises in its rate parameter (which is inversely proportional to its mean), we performed the RPCA after taking the entry-wise inverse of the data matrices. Fig. 3b shows the reconstruction errors for **L** and **S** (in Frobenius norm) using the $e^{\text{RPCA}}$, the standard RPCA and the ePCA, where again the latter does not provide an estimate of **S**. As before, the $e^{\text{RPCA}}$ yields improved recovery of both the low-rank structure **L** and the sparse anomalies **S**, with this improvement gap growing as dimension $p$ increases. Figs. 5a and 5b visualize the recovery of **L** and **S**, respectively, for a simulation experiment in $p = 10$ dimensions. The standard RPCA, which does not factor for non-Gaussian noise, can be seen to yield poorer recovery of **L** and erroneous detection of numerous false "anomalies". The ePCA offers a slightly better recovery of **L**, but this is highly corrupted by the underlying sparse outliers. By integrating both non-Gaussian noise and sparse anomalies within an efficient decomposition framework, the $e^{\text{RPCA}}$ facilitates an accurate recovery of both **L** and **S**.

*3) Poisson Distribution:* Finally, consider noise drawn from the Poisson distribution, where each matrix entry follows $M_{i,j,k} \overset{indep.}{\sim} \text{Pois}(\theta_{j,k})$. Fig. 3c shows the recovery errors for **L** and **S** using the compared methods. Here, we see that the $e^{\text{RPCA}}$ and standard RPCA have comparable performance, with both having considerably lower errors than the ePCA. In particular, $e^{\text{RPCA}}$ yields slightly lower errors for both **L** and **S** in lower dimensions, while the standard RPCA performs slightly better in higher dimensions. One likely reason for this is that, with sufficiently large rate parameters $\theta$, the resulting Poisson noise can be well-approximated by a Gaussian distribution [17]. Thus, for Poisson noise, both the $e^{\text{RPCA}}$ and the standard RPCA yield good performance. Figs. 6a and 6b visualize the recovered **L** and **S** for an experiment in $p = 10$ dimensions. We see that the $e^{\text{RPCA}}$ and the standard RPCA both provide good recovery of both the low-rank structure and sparse anomalies, with the standard RPCA again erroneously

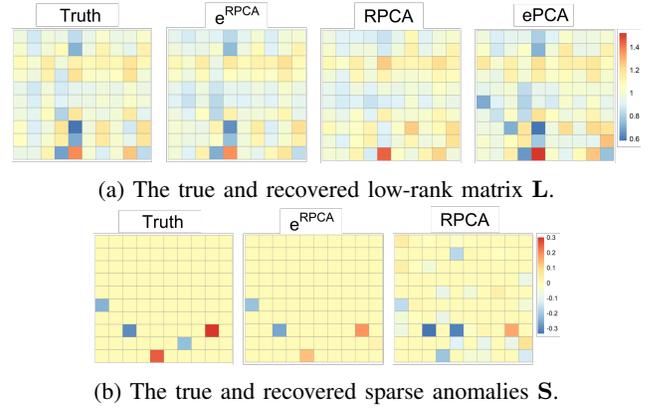

(a) The true and recovered low-rank matrix **L**.

(b) The true and recovered sparse anomalies **S**.

Fig. 5: Visualizing the true and recovered low-rank matrix **L** and sparse anomalies **S** in one simulation in $p = 10$ dimensions, for the single-group Exponential simulations.

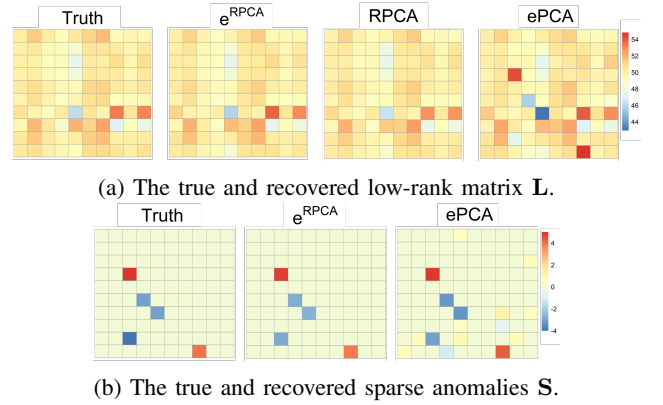

(a) The true and recovered low-rank matrix **L**.

(b) The true and recovered sparse anomalies **S**.

Fig. 6: Visualizing the true and recovered low-rank matrix **L** and sparse anomalies **S** in one simulation in $p = 10$ dimensions, for the single-group Poisson simulations.

identifying more false "anomalies".

### B. Multi-Group Setting

Next, we performed multi-group experiments with $G = 2$ groups. We generated the low-rank matrix **L** and sparse anomalies $\mathbf{S}_1$ and $\mathbf{S}_2$ as before, then simulated $n_1 = n_2 = 250$ data matrices for each group, with $\Theta_1 = \mathbf{L} + \mathbf{S}_1$ and $\Theta_2 = \mathbf{L} + \mathbf{S}_2$. We then compared methods on its recovery of the low-rank structure **L** and the sparse anomalies **S** in terms of Frobenius error. These experiments are again performed for $p \times p$ matrices, where $p = 10, 20, 30$ and $40$, with each setting replicated for 30 trials.

*1) Bernoulli Distribution:* Consider first the multi-group setting with Bernoulli noise. Fig. 7a shows the recovery errors for **L** and **S** using the $e^{\text{RPCA}}$, the RPCA, and the ePCA, where again the latter does not provide an estimate for **S**. We see that, for **L**, the $e^{\text{RPCA}}$ offers improved recovery to competing methods, with the improvement growing larger as dimension $p$ increases. Furthermore, the $e^{\text{RPCA}}$ appears to yield greater improvement in this multi-group experiment compared to the earlier single-group experiment. This is not too surprising, as these competing methods do not factor in the different sparse anomalies between different groups. Similar observations can be made for the recovery of sparse anomalies **S**.

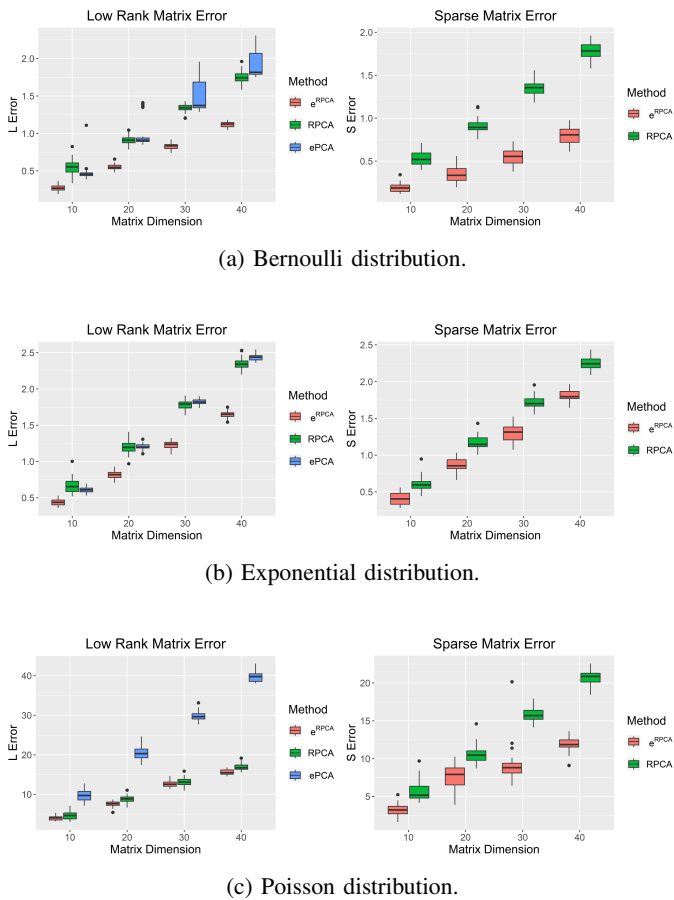

Fig. 7: Boxplots of recovery errors (in Frobenius norm) for the low-rank matrix **L** (left), and sparse anomalies **S** (right) as a function of matrix dimension $p$ for multi-group simulations across different distributions.

*2) Exponential Distribution:* Consider next the multi-group setting with Exponential noise. Fig. 7b shows the recovery errors for **L** and **S** using $e^{\text{RPCA}}$ and competing methods. For **L**, we again see a marked improvement for the proposed method, with this improvement growing as dimension $p$ increases. As before, these improvements appear more pronounced for the multi-group set-up compared to the single-group set-up. Similar observations hold for the recovery of anomalies in **S**.

*3) Poisson Distribution:* Finally, consider the multi-group setting with Poisson noise. Fig. 7c shows the recovery errors for **L** and **S**. Interestingly, while the earlier single-group experiments showed comparable results for the $e^{\text{RPCA}}$ and RPCA, the multi-group experiments show a noticeable improvement for the $e^{\text{RPCA}}$. This can again be explained by the presence of different sparse anomalies within different groups, which is not accounted for in the standard RPCA. Similar conclusions hold for recovering the sparse anomalies **S**.

## V. APPLICATIONS

We now explore the use of the proposed $e^{\text{RPCA}}$ in two practical applications. The first is our motivating problem on steel defect detection, and the second is a crime monitoring application in the city of Atlanta.

### A. Steel Defect Detection

Consider first the motivating steel defect detection problem from Section II-C. Steel manufacturing is essential in many facets of modern manufacturing, including the production of automobiles, electronics, furniture, infrastructure, and shipbuilding. The automated monitoring of steel defects, e.g., gashes, dents, or other inhomogeneities, thus plays a critical role in maintaining high product quality at low operation costs. Defects are typically detected via careful monitoring of images of the steel sheets taken from high-frequency cameras. Recent studies, e.g., [9], have shown that quanta image sensing (QIS, [11]) may offer improved higher frequency imaging with lower read noise over more conventional multi-bit imaging systems (e.g., complementary metal-oxide semiconductor imaging [11]). The key challenge in defect detection using QIS is that its image intensities take the form of binary bits, which can be well-modeled via i.i.d. Bernoulli noise [9]; this thus presents an appropriate application for the $e^{\text{RPCA}}$.

We adopt the same set-up as Section II-C for numerical experiments. First, an uncorrupted steel sheet image with visible defects is taken from Severstal [34] (see Fig. 1 left). Next, to mimic QIS, $n = 500$ binary images are generated by normalizing the uncorrupted image and sampling via i.i.d. Bernoulli noise (see Fig. 1 right). The compared methods include the proposed (single-group) $e^{\text{RPCA}}$, the standard RPCA (with modifications as discussed in Section IV), and the ePCA. For the $e^{\text{RPCA}}$, we made use of the default hyperparameter specification in Section III-C, i.e., without any prior knowledge of rank or degree of sparsity.

Fig. 8a shows the recovered low-rank structure **L** using the standard RPCA, ePCA, and the proposed $e^{\text{RPCA}}$. Here, the desired structure to recover is the underlying "criss-cross" background pattern. As observed previously, the first two existing methods yield a visually mediocre recovery of this criss-cross pattern; one reason may be that neither method accounts for the *joint* presence of sparse anomalies with non-Gaussian noise. The $e^{\text{RPCA}}$, by factoring in both properties, in turn, provides a noticeably smoother recovery of the cross-cross pattern without defects.

Fig. 8b shows the recovered sparse anomalies **S** using RPCA and $e^{\text{RPCA}}$; note that the ePCA does not provide an estimate of **S**. From Fig. 1, the desired anomalies to recover include a large gash on the right and two smaller inhomogeneous spots in the middle. We see that the RPCA returns a rather muddled recovery of such defects: while the large right gash is noticeable visually, the recovered **S** also picks up on the background criss-cross structure, which obfuscates other defects. Comparatively, the $e^{\text{RPCA}}$ yields improved recovery of the underlying defects: it picks up not only the clear right gash and the more subtle inhomogeneities in the middle but also darker discolorations along the criss-cross structure on

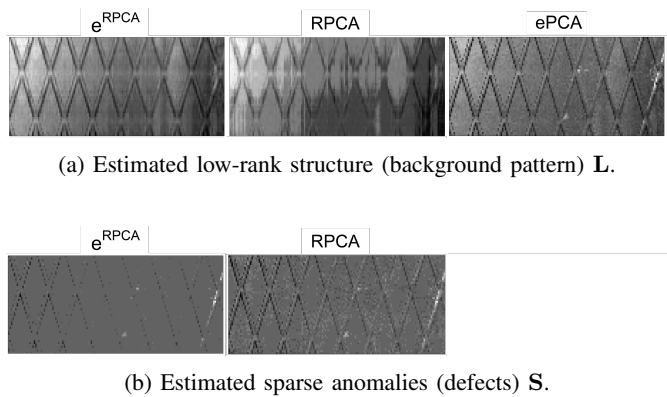

(a) Estimated low-rank structure (background pattern) **L**.

(b) Estimated sparse anomalies (defects) **S**.

Fig. 8: Visualizing (a) the estimated low-rank structure (background pattern) **L** and (b) the recovered sparse anomalies (defects) **S**, using the $e^{\text{RPCA}}$, RPCA and ePCA in the steel defect detection application.

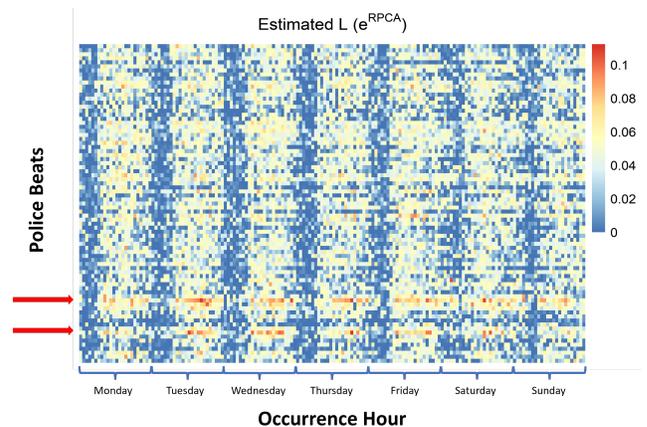

Fig. 9: Visualizing the estimated low-rank matrix **L** from $e^{\text{RPCA}}$ for the crime monitoring application. The two red arrows highlight two police beats: Beat 512 (downtown Atlanta) and Beat 509 (Midtown Atlanta).

the left. The latter defects were not immediately evident at first glance from Fig. 1, but are indeed present upon further inspection from this analysis. Thus, by integrating information on non-Gaussian noise, the proposed $e^{\text{RPCA}}$ appears capable of jointly recovering low-rank structures and identifying sparse defects for steel monitoring.

### B. Crime Mapping in Atlanta

Next, we investigate the use of the $e^{\text{RPCA}}$ for a crime monitoring application in the city of Atlanta. The identification of *regular* (background) and *irregular* (anomalous) criminal activities are clearly critical tasks for improving public safety: it helps inform law enforcement of hotspots and/or peak times for different crimes [14], and facilitates the diagnosis of abnormal crime spikes. As such, a key objective here is to identify regular and irregular crime patterns over different geographical regions and times.

Our data consists of reported burglaries from the Atlanta Police Department over two years: 2015 and 2016. Each burglary is recorded along with its hour of occurrence, as well as its spatial location in the form of police "beats" (or zones) for patrol, of which there are 79 beats in total. Given the fine spatiotemporal scales for the recorded burglaries, there is typically at most one recorded crime within each spatiotemporal "window", i.e., a combination of occurrence hour and police beat. Prior studies [19, 32] suggest that *weekly* recurrent trends may be common for burglaries, which we leverage next for specifying the underlying low-rank structure.

Since there is no ground truth available for **L** and **S** here, we forgo a full comparison with existing methods and instead investigate the extraction of useful burglary patterns (both regular and irregular) from the proposed $e^{\text{RPCA}}$. Since there is at most one recorded crime within nearly all spatiotemporal windows, we thus adopt a binarization of this data ("0" for no burglaries within the window, "1" for at least one burglary). With this, $e^{\text{RPCA}}$ then proceeds using the Bernoulli distribution, with the underlying probability matrix $\Theta$ presumed to follow the low-rank plus sparse decomposition. We employ here the multi-group $e^{\text{RPCA}}$ set-up, with $G = 2$ groups for the summer and winter seasons (more on this next). To incorporate information on weekly trends, the data matrices $\{\mathbf{M}_i^{[g]}\}_{i=1}^{n_g}$ are constructed week-by-week; each matrix is thus of dimensions $79 \times 168$, where $p = 79$ is the number of police beats, and $q = 24 \times 7 = 168$ is the number of hours in a week. We then take $n_g = 24$ weeks (12 weeks per season $\times$ 2 years) of reported crime data for both the summer and winter seasons. With this set-up, the low-rank matrix **L** models for weekly (regular) crime activity, and the sparse matrices $\mathbf{S}_1$ and $\mathbf{S}_2$ account for irregular burglary activity over the summer and winter seasons, respectively.

Fig. 9 shows the recovered low-rank matrix **L** using the two-group $e^{\text{RPCA}}$. We immediately see seven bright vertical bands, which suggest the presence of a daily trend in crime activity. In particular, for most beats, we observe an increased probability of burglary after dawn, which peaks during the day and decreases during the evening. Several beats, such as Beat 512 (downtown Atlanta) or 509 (Midtown Atlanta), have noticeably higher burglary rates compared to other beats, which is expected as such areas are highly urban and dense in terms of population. There also seems to be some interactions between occurrence hour and police beat, in that some beats have notably different peak hours than other beats.

Fig. 10 visualizes the spatial distribution of the estimated low-rank matrix **L** at various hours in the week. Certain beats, such as Beat 202 and 209 (outlined in blue), have noticeably higher burglary rates on Monday mornings, whereas other beats, such as Beat 203 or 313 (outlined in green), experience higher rates in the evenings. Upon further inspection, this is quite intuitive: the former beats (in blue) are primarily residential areas, where burglaries are expected to occur more often during weekday mornings (e.g., when individuals are at work), whereas the latter beats (in green) are primarily business districts, where burglaries typically occur more frequently at night (e.g. when businesses are closed). The estimated **L** reflects such intuitive patterns, thus suggesting the proposed $e^{\text{RPCA}}$ can indeed recover useful insights for diagnosis.

Next, Fig. 11 visualizes the estimated sparse matrices $\mathbf{S}_1$

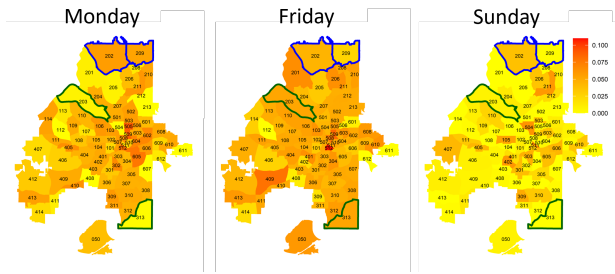

(a) Spatial visualization of burglary rates at 9am.

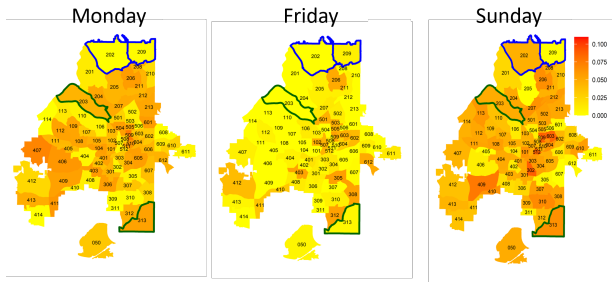

(b) Spatial visualization of burglary rates at 10pm.

Fig. 10: Spatial visualization of the burglary rate from the estimated $\mathbf{L}$ on Monday, Friday, and Sunday at (a) 9 a.m. and (b) 10 p.m. for the crime monitoring application. Beats 202 and 209 (outlined in blue) have noticeably higher rates on Monday mornings, whereas Beats 203 or 313 (outlined in green) experience higher rates in the evenings.

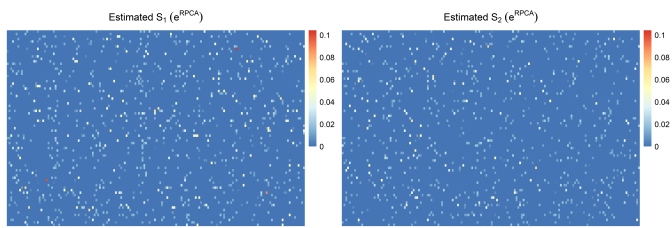

Fig. 11: Visualizing the absolute values of the estimated anomaly matrices $\mathbf{S}_1$ (summer) and $\mathbf{S}_2$ (winter) from $e^{\text{RPCA}}$ for the crime monitoring application.

and $\mathbf{S}_2$, which models for anomalous (irregular) burglary activities over the summer and winter seasons, respectively. From a quick inspection, there appear to be some differences in anomalies between the two seasons, which supports our multi-group approach. To explore this difference, Fig. 12 plots the non-zero entries in $\mathbf{S}_1$ and $\mathbf{S}_2$ over Beats 403 (a residential area) and 503 (a commercial area). For Beat 403, it appears that after factoring in the weekly trends in $\mathbf{L}$, it experiences more anomalous burglaries in the summer than in the winter during the evenings. A plausible explanation is that, in residential areas, burglaries during the daytime become more difficult due to more people being at home from summer break or hot weather. Similarly, for Beat 503, one observes a greater quantity of irregular burglaries in the summer than in the winter. A likely reason is that commercial areas typically experience a large influx of tourists and visitors, which may create greater opportunities for burglaries. Such insights suggest that the proposed $e^{\text{RPCA}}$ can indeed extract interpretable and useful insights for crime monitoring.

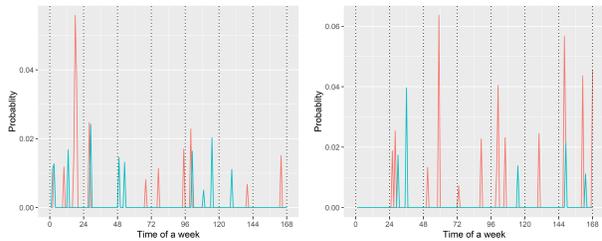

Fig. 12: Plotting the non-zero entries in $\mathbf{S}_1$ (summer) and $\mathbf{S}_2$ (winter) for Beat 403 (left, residential) and Beat 503 (right, commercial) from $e^{\text{RPCA}}$ for the crime monitoring application.

## VI. CONCLUSION

In this work, we proposed a new $e^{\text{RPCA}}$ method for jointly recovering embedded low-rank structures and corresponding sparse anomalies from data matrices corrupted by non-Gaussian noise from the exponential family distribution. This method is directly motivated by two applications, the first for steel defect detection and the second for crime activity monitoring. The $e^{\text{RPCA}}$ employs a novel optimization formulation that leverages the underlying exponential noise structure for performing the desired low-rank plus sparse decomposition. We then presented an alternating direction method of multiplier algorithm for efficient optimization, both for the single-group $e^{\text{RPCA}}$ (where anomalies are shared over all data matrices) and the multi-group $e^{\text{RPCA}}$ (where anomalies may vary over different groups). We demonstrated the effectiveness of the proposed $e^{\text{RPCA}}$ in a suite of numerical experiments with varying non-Gaussian noise and in our two motivating applications for steel defect detection and crime monitoring.

Given promising results, there are numerous directions for impactful future work. First, a Bayesian extension of the $e^{\text{RPCA}}$ would be of interest, as in many applications, it would be useful to have a reliable quantification of uncertainty for anomaly detection. Recent work on Bayesian matrix modeling [41] appears promising on this front. Another direction is the application of such methods for high-energy physics [15, 23, 26], where there has been much recent work on using anomaly detection techniques to identify new particle activity from heavy-ion collisions; see, e.g., [24]. A key challenge is the non-Gaussian measurement noise in such systems [18], and the $e^{\text{RPCA}}$ can be highly useful in this setting. The decoupling of shared and unique features from diverse data sources (see, e.g., [37]) is also of potential interest as future work.

ACKNOWLEDGEMENT

XZ and SM gratefully acknowledge support from NSF CSSI 2004571, NSF DMS 2210729 and DE-SC0024477. LX is supported by UDF01002142 and 2023SC0019 through the Chinese University of Hong Kong, Shenzhen. YX is partially supported by an NSF CAREER CCF-1650913, NSF DMS-2134037, CMMI-2015787, CMMI-2112533, DMS-1938106, DMS-1830210, and the Coca-Cola Foundation.

## APPENDIX

### A. Closed-form Optimization Updates for $\theta$

### B. Simulation Set-up for $\mathbf{L}$ and $\mathbf{S}$

The settings for constructing the low-rank matrix $\mathbf{L}$ are as follows. For the Bernoulli distribution, we use $\mu = 0.5, \sigma = 0.15$. For the Exponential distribution, $\mu = 1, \sigma = 0.15$; and for the Poisson distribution, $\mu = 50, \sigma = 2$. When creating the sparse matrix $\mathbf{S}$, the spike ranges are $[0.2, 0.3]$ for Bernoulli distribution and Exponential distributions, and $[2, 5]$ for the Poisson distribution. Note that the Bernoulli parameters must lie between $[0, 1]$, so any out-of-range sampled values are set to zero or one (whichever is closest). For the Exponential and

| Distribution | $\theta_{j,k}^*$ |
|---|---|
| Poisson | root of $a\theta_{j,k}^2 + b\theta_{j,k} + c = 0$ with the smallest loss, where $a = \mu,$ $b = -\mu(L_{j,k} + S_{j,k}) + Y_{j,k} + 1,$ $c = -\bar{M}_{j,k}$ |
| Bernoulli | root of $a\theta_{j,k}^3 + b\theta_{j,k}^2 + c\theta_{j,k} + d = 0$ with the smallest loss, where $a = -\mu,$ $b = \mu(1 + L_{j,k} + S_{j,k}) - Y_{j,k},$ $c = 1 - \mu(L_{j,k} + S_{j,k}) + Y_{j,k},$ $d = -\bar{M}_{j,k}$ |
| Exponential | root of $a\theta_{j,k}^2 + b\theta_{j,k} + c = 0$ with the smallest loss, where $a = \mu,$ $b = -\mu(L_{j,k} + S_{j,k}) + Y_{j,k} + \bar{M}_{j,k},$ $c = -1$ |
| Gaussian | $(\bar{M}_{j,k} + \mu(L_{j,k} + S_{j,k}) - Y_{j,k})/(1 + \mu)$ |

TABLE II: Closed-form updates of (19) for common (one-parameter) exponential family distributions.

Poisson distributions, only positive parameters are permitted, thus negative sampled values are set as zero.